# HIGH POWER TESTING RESULTS OF THE X-BAND MIXED-MODE RF WINDOWS FOR LINEAR COLLIDERS

S. Tokumoto, Y. H. Chin, H. Mizuno, K. Ohya, S. Yamaguchi , KEK, Tsukuba, Japan
S. Kazakov, BINP, Protvino, Moscow, Russia
R. Loewen, R. Fowkes, A. Menegat, A. Vlieks, SLAC, Stanford, CA, USA

*Abstract*

In this paper, we summarize the high power testing results of the X-band mixed-mode RF windows at KEK and SLAC for linear colliders. The main feature of these windows is that the combination of modes on the surface of the ceramic significantly decreases the electric and magnetic fields in the junction between the ceramic and the metal. So far two types of high power windows (with the diameter of 53 mm and 64 mm) have been fabricated. A high power model of the smaller type window was fabricated and tested in a resonant ring at KEK. A maximum circulating power of 81 MW with 300 ns duration or 66 MW with 700 ns duration was achieved. Light emission was observed for a power level of over 10 MW. Later, both windows were shipped to SLAC for even higher power testing using combined power from two klystrons. The first window (53 mm diameter) achieved a transmitted power of 80 MW with 1.5 microsec duration at 30 Hz repetition. It was not destroyed during the testing. The testing of the second window was carried out next to the small type and achieved 62 MW with 1.5 maicrosec duration with 10 Hz repetition. The final results of both windows are presented in this report.

## 1 INTRODUCTION

In the future linear collider (JLC) project [1], about 3300 of X-band klystrons will be used. An output power of 75 MW with 1.5 µs duration is required per klystron. For handling such high output power with high reliability or for keeping the klystron a long lifetime, a RF window is one of the key technologies. Therefore some types of RF windows through which high power RF pulses of around 100 MW is capable to transmit have been developed among these R&D works.

Destruction of the RF window is caused by high frequency electrical breakdown at the surface of the window ceramic due to high electric field there. As was shown in Ref. [2], it is considered the maximum field strength is 8 kV/mm with highly purified $Al_2O_3$. One of the effective electrical designs to decrease electric field strength on the ceramic surface down under the threshold is to increase the diameter of the ceramic up to large enough. Some RF windows based on this idea have been developed and tested in KEK [3]. In those tests, it was found that there was a weak point near the periphery of the ceramic, where some tracks of breakdown were left along the direction of the electric field of the transmitted mode.

Taking attention with those, an X-band mixed-mode RF window with low electric field at the ceramic, especially in ceramic-metal brazing area, was newly designed. The principle of decreasing electric field strength is to superpose a couple of electric fields eachother. They are produced by two different modes propagating through the ceramic at the same time and have an opposite field direction. Another point is to use the structure where traveling wave is transmitted through the ceramic. Details of designing were described in another paper [4]. Two types of X-band mixed mode windows were fabricated and tested at KEK and SLAC. In this paper, results of high power testing are presented.

## 2 MIXED-MODE RF WINDOW

The X-band RF windows developed before was designed to use only one mode, which was TE11. Any of them were designed to use over-sized window ceramic where electric field strength on the surface was relatively decreased compared to that of the rectangular waveguide. They have been sustained with the transmitted RF power of more than 100 MW. The mixed-mode window was designed referring to their structure. A schematic cross section of the X-band mixed mode window is illustrated in Fig. 1.

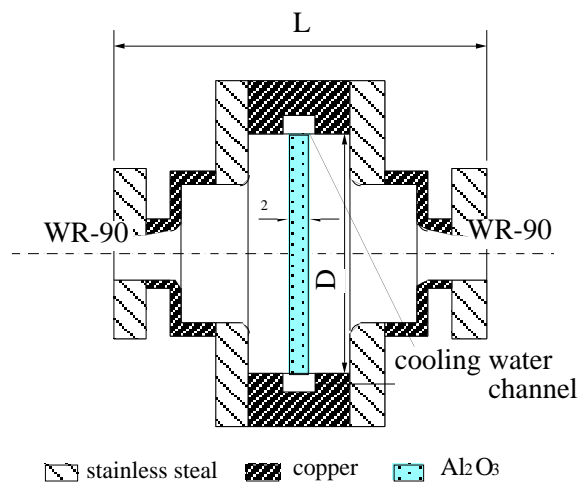

Fig.1: Schematic cross-section of the mixed mode RF window. Length (L) is 85 mm for small size (D=53mm) and 115mm for large size (D=63 mm)

Two types of the window of which the ceramic are 53 mm in diameter and 2mm in thickness, and 64 mm and 2.15 mm were fabricated. The thickness of the ceramic was chosen to be close to 1/4-wavelength to tune away the operating frequency from the "ghost" mode. Both of the ceramic was treated on their surface with TiN coating in order to suppress multipactoring. Coating thickness was chosen to 10 nm. Combinations of the mode propagating through the ceramic were chosen TE11-mode plus TM11-mode or TE11-mode plus TE12-mode. Consequently it determined a diameter of the ceramic as TE11+TM11 for small size and TE11+TE12 for large. Measured bandwidth at –20 dB was more than 300 MHz.

The electric field strength on the ceramic was computed with using the HFSS Code. According computed results, when a high RF power of 100 MW is transmitted through the window, maximum field strength on the ceramic surface such as E(normal), E(tangential) and E(periphery) were predicted as shown in table 1. For comparing these results to those of other windows designed to make low electric field on the ceramic surface, some computed and tested results of successively tested windows in KEK are also shown in table 1.

Table 1. Comparison field strength of windows successively tested at KEK

|  | E [MV/m] @ 100MW (HFSS) | | | $P_{tested}$ [MW] (Pulse Width) |
|---|---|---|---|---|
|  | $E_{tan}$ | $E_{nor}$ | $E_{per}$ |  |
| PILLBOX | 16 | 19 | 7 | 30 (200ns)* |
| TE11-MODE WITH TAPER TRANSITION | 8 | - | 7 | 130 (300ns) / 84 (700ns) |
| TE11-MODE WITH TRAVELING-WAVE | 7 | - | 7 | 90 (200ns) |
| MIXED-MODE TE11 + TM11 | 7 | 6 | 1 |  |
| MIXED-MODE TE11 + TE12 | 6 | 6 | 1 |  |

*Tested with kly. Output

## 3 HIGH POWER TEST

High power tests of the windows were carried out at KEK and SLAC. At first a small type window was tested with using an X-band resonant ring at KEK. Next, both two types of the window were tested with higher and longer RF pulse at SLAC.

### 3.1 TEST AT KEK

An X-band resonant ring comprised a 10.5 dB directional coupler, a couple of three stub tuner, a phase shifter, a couple of pumping ports attached 30 l/s ion pumps and a 60 dB directional coupler for power monitoring. The high power test was carried out for about 80 hours and maximum circulating power of 81 MW with input pulse length of 300 ns or 66 MW with 700 ns was achieved. During the test, vacuum pressure was about 2 x $10^{-5}$ Pa with RF pulse. Light emission on the surface was observed with TV camera and photo multiplier tube (PMT) through viewing ports at both side of the window. PMT signals appeared over 10MW and rose up 10 times as large at 50 MW. Faint dark blue light was observed with camera. Around 80 MW with short pulse length or 60 MW with longer pulse, sudden rise up of vacuum pressure were frequently occurred and system was tripped. At that time, PMT signal was over-scaled and left its trace long time after RF pulse was stopped. Bright blue or white light was also observed on the surface. The processing history is illustrated in Fig.2

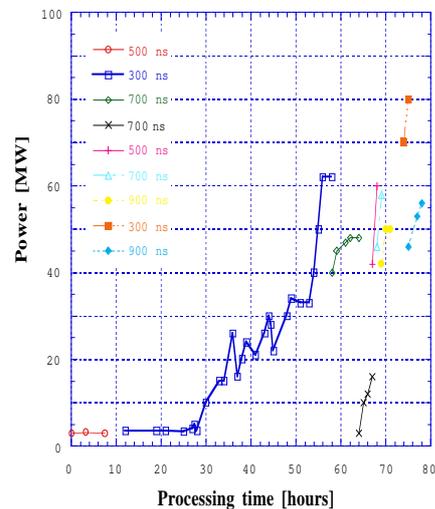

Fig. 2: Processing history for the small type window at KEK

### 3-2 TEST AT SLAC

At first, a small type of the window was tested for about 100 hours after 72 hours baking at 120 degrees. Transmitted RF power rose up smoothly compared to KEK. With long pulse length of 1.5 microsec, breakdown occurred frequently and vacuum pressure suddenly rose up with flashover on the surface. The maximum transmitted power of 80 MW with 1.5 microsec duration was finally achieved. Fig.3 shows a processing history for small type window.

Next to the small type window, the large type window was tested for about 100 hours. This window was also treated a baking at 120 degrees for 72 hours. However, it took a little more time to rose up RF power and with 1.5 microsec pulse, breakdown with big out-gassing were frequently occurred. The maximum transmitted power of 62 MW with 1.5 microsec duration was finally achieved. Fig.4 shows a processing history for the large type window.

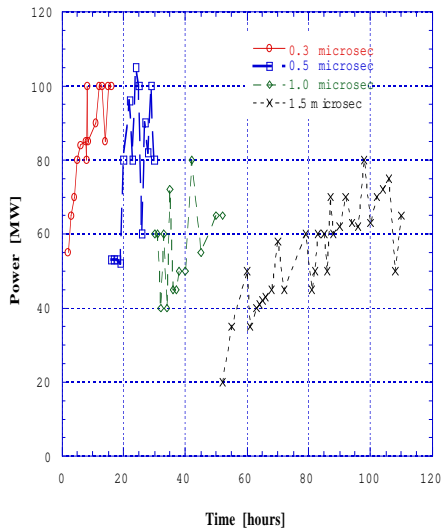

Fig.3: Processing history for the small type window at SLAC

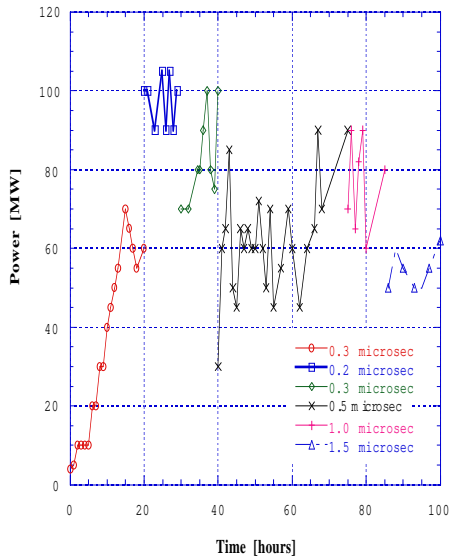

Fig.4: Processing history for the large type window

During high power tests of both windows, light emission on the surface was observed with PMT and TV camera. With shorter pulse of around 0.5 microsec, faint dark blue light was observed around 50 MW and more than 50 MW it has disappeared. At that time, PMT signal went smaller with decreasing of vacuum pressure. With longer pulse, flashover with bright blue light on the surface was frequently observed over 50 MW. PMT signal showed that increasing of light emission depended on a duration and power level of transmitted pulse.

Observed PMT signals at transmitted power between 50 MW and 68 MW are shown in Fig.5

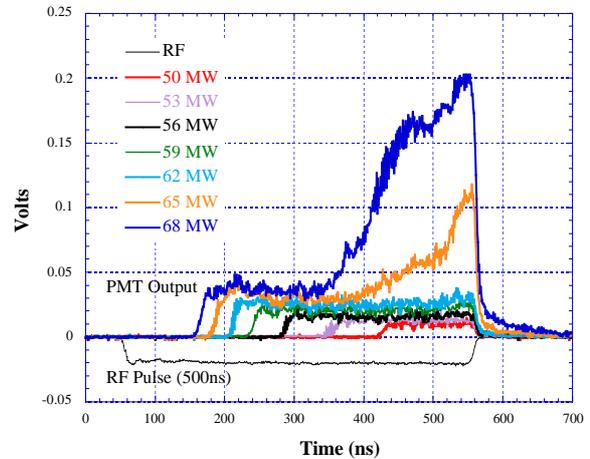

Fig. 5: Observed PMT signals at high power test of large size window.

## SUMMARY

We summarize high power testing results:
- Both two types of the X-band mixed-mode windows showed a good capability at testing with high RF power of over 100 MW.
- During all the testing, there was no window destructed.
- Combination of modes effectively decreases electric field strength at periphery so as to avoid serious breakdown at the brazing area.